\documentclass[%
reprint,
superscriptaddress,
preprintnumbers,
amsmath,amssymb,
aps,
prd,
]{revtex4-2}

\usepackage[utf8]{inputenc}

\usepackage{mathtools}
\usepackage{amsfonts}
\usepackage{mathrsfs}
\usepackage{bm}
\usepackage{bbold}
\usepackage{slashed}
\usepackage{dsfont}
\usepackage{float}
\usepackage{dcolumn}
\usepackage{amssymb,amsmath}

\usepackage{graphicx}
\usepackage{subcaption}
\usepackage{color}
\usepackage{array}

\usepackage{placeins}
\usepackage{booktabs}
\usepackage{caption}

\usepackage{xspace}
\usepackage{hyperref}
\usepackage[nameinlink]{cleveref}
\usepackage{bookmark}
\usepackage{units}

\def\Eq#1{Eq.~\labelcref{#1}}

\def\Fig#1{Fig.~\labelcref{#1}}
\def\Tab#1{Tab.~\labelcref{#1}}
\def\sec#1{Sec.~\labelcref{#1}}

\setkeys{Gin}{width=0.48\textwidth}
\usepackage{booktabs}
\usepackage{multirow}

\newcolumntype{C}{>{$}c<{$}}
\AtBeginDocument{
	\heavyrulewidth=.08em
	\lightrulewidth=.05em
	\cmidrulewidth=.03em
	\belowrulesep=.65ex
	\belowbottomsep=0pt
	\aboverulesep=.4ex
	\abovetopsep=0pt
	\cmidrulesep=\doublerulesep
	\cmidrulekern=.5em
	\defaultaddspace=.5em
}

\graphicspath{{./figures/}}

\usepackage{xifthen}
\usepackage{xcolor}

\newcommand{\gettitle}{Criticality of the $O(N)$ universality via global solutions to nonperturbative fixed-point equations}

\hypersetup{
	colorlinks,
	linkcolor={red!75!black},
	citecolor={blue!75!black},
	urlcolor={blue!75!black}, 
	pdftitle={\gettitle},
	pdfauthor={},
	pdfkeywords={}
	{} 
	bookmarksopen=true,
	bookmarksopenlevel=2,
	bookmarksnumbered=true
}

\begin{document}
\title{\gettitle}
	
\author{Yang-yang Tan}
\affiliation{School of Physics, Dalian University of Technology, Dalian, 116024, P.R. China}	

\author{Chuang Huang}
\affiliation{School of Physics, Dalian University of Technology, Dalian, 116024, P.R. China}	

\author{Yong-rui Chen}
\affiliation{School of Physics, Dalian University of Technology, Dalian, 116024, P.R. China}	

\author{Wei-jie Fu}
\email{wjfu@dlut.edu.cn}
\affiliation{School of Physics, Dalian University of Technology, Dalian, 116024, P.R. China}

\begin{abstract}

Fixed-point equations in the functional renormalization group approach are integrated from large to vanishing field, where an asymptotic potential in the limit of large field is implemented as initial conditions. This approach allows us to obtain a global fixed-point potential with high numerical accuracy, that incorporates the correct asymptotic behavior in the limit of large field. Our calculated global potential is in good agreement with the Taylor expansion in the region of small field, and it also coincides with the Laurent expansion in the regime of large field. Laurent expansion of the potential in the limit of large field for general case, that the spatial dimension $d$ is a continuous variable in the range $2\leq d \leq 4$, is obtained. Eigenfunctions and eigenvalues of perturbations near the Wilson-Fisher fixed point are computed with the method of eigenperturbations. Critical exponents for different values of $d$ and $N$ of the $O(N)$ universality class are calculated.

\end{abstract}

\maketitle

\section{Introduction}
\label{sec:int}

The renormalization group (RG) has been recognized as a powerful theoretical tool to study phase transitions and critical phenomena since the seminal work by Wilson {\it et al.} \cite{Wilson:1971bg, Wilson:1971dh, Wilson:1971dc, Wilson:1973jj}. A second-order phase transition is regarded to be located on the critical surface of a stable fixed point of RG flow equations. Critical behaviors of the phase transition, such as universal critical exponents and their relations to the dimension and symmetries, are closely connected to relevant eigenperturbations in the proximity of the fixed point, see, e.g., \cite{Ma:2020a} for more details.

When the spatial dimension $d$ is smaller than and close to 4, that is, the parameter $\varepsilon \equiv 4-d$ being a positive and small quantity, the RG flows can be expanded in powers of $\varepsilon$ \cite{Wilson:1971dc, Wilson:1973jj}. In other words, one is able to use the technique of perturbation theory to compute, e.g., critical exponents, in powers of $\varepsilon$ order by order. Nevertheless, the reliability of perturbation theory is gradually loosened with the increase of the expansion parameter $\varepsilon$, e.g., in the case that the dimension is $d=3$, or even approaches toward 2. In such cases, nonperturbative RG flows are indispensable. 

The functional renormalization group (fRG) provides us with a convenient framework to deal with nonperturbative RG flows \cite{Wetterich:1992yh}. In the fRG approach physics of nonperturbation theory are encoded in a self-consistent flow equation for the effective action or effective potential. For more details about the method of fRG and its applications in studies of nonperturbative physics, such as the strongly correlated QCD, one is referred to, e.g., \cite{Berges:2000ew, Pawlowski:2005xe, Braun:2011pp, Dupuis:2020fhh, Fu:2022gou} for reviews and \cite{Braun:2014ata, Mitter:2014wpa, Rennecke:2015eba, Cyrol:2016tym, Cyrol:2017qkl, Cyrol:2017ewj, Fu:2019hdw, Braun:2020ada, Fu:2022uow} for recent progresses in first-principle fRG calculations in QCD.

Truncations for the RG flows in the fRG approach can be usually made in a systematic way, such as the derivative expansions (DE) \cite{Balog:2019rrg, DePolsi:2020pjk}, which provides us with a set of closed flow equations or fixed-point equations for the effective potential and other dressing functions. In order to obtain scaling solutions of RG flows, one can evolve the flow equations into a scale-independent solution \cite{Bohr:2000gp,Papp:1999he}, or directly solve the fixed-point equations. To solve, e.g., the fixed-point equation for the potential, one can either use the Taylor expansion to expand the potential around vanishing or finite field if the dimension $d$ is not too small \cite{Litim:2002cf}, or integrate the fixed-point equation starting from the vanishing field \cite{Codello:2012sc}. It is found that direct integration of the fixed-point equation starting from the vanishing field would end up in a singularity at a finite field \cite{Codello:2012sc}, which implies that a proper treatment at large field is necessary. Studies of the global solution to the fixed-point equation with the correct asymptotic behavior in the limit of large field in the fRG approach have made progress in recent years. A combined technique with both small and large field expansion is used to obtain the global potential of the fixed point \cite{Juttner:2017cpr}. Moreover, pseudospectral methods are also employed to construct global solutions \cite{Borchardt:2015rxa} or to evolve the flow equation \cite{Chen:2021iuo}. Recently, discontinuous Galerkin methods have been developed to resolve the flow equation of a global potential \cite{Grossi:2019urj, Grossi:2021ksl, Ihssen:2022xkr}.

In calculating the global potential to a fixed-point equation, the whole range of the argument of potential, i.e., the field, is usually segmented into several subranges, e.g. the region of large field and that of small field. Different methods are applied in different regimes, such as, Taylor expansion in the region of small field and numerical calculations in that of large field \cite{Juttner:2017cpr}, a standard Chebyshev series in small field and a rational Chebyshev series in large field \cite{Borchardt:2015rxa}. A global potential is finally obtained by connecting the solutions in different regimes. In this work, we try to simplify the procedure, and would like to directly integrate the fixed-point equation starting at a sufficient large value of field, where the asymptotic potential in the limit of large field is implemented as initial conditions. As a consequence, a global fixed-point potential with high numerical accuracy is resolved after the integration is finished at vanishing field, that naturally incorporates the correct asymptotic behaviors both in the limit of vanishing and in the large field. Moreover, we would like to discuss the Laurent expansion of the potential in the limit of large field for a general case  that the spatial dimension $d$ is a continuous variable in the range $2\leq d \leq 4$.

This paper is organized as follows: In \sec{sec:flow} we give a brief introduction about the fRG approach with the flow equation of the effective potential and the fixed-point equation, as well as some notations used thereafter. The Taylor expansion of the potential around vanishing or finite field and the Laurent expansion in the limit of large field are discussed in \sec{sec:local-solu}. In \sec{sec:global-solu} we discuss how the fixed-point equation is integrated out with the asymptotic potential in the limit of large field implemented. Moreover, eigenperturbations near the fixed point are also discussed. In \sec{sec:num-resul} we present numerical results on fixed-point potentials and critical exponents. Finally, a summary with outlook is given in \sec{sec:summary}.

\section{Flow equation of the effective potential}
\label{sec:flow}

We begin with a RG scale $k$-dependent effective action for the $O(N)$ scalar theory, which reads
\begin{align}
   \Gamma_{k}[\phi]=&\int \mathrm{d}^d x \left[\frac{1}{2}Z_{\phi,k}\left(\partial_{\mu} \phi_{a}\right)^{2}+V_{k}(\rho)\right]\,,\label{eq:action}
\end{align}
with $\mu=1,\,2,\cdots d$ and $a=0,\,1,\,2,\cdots N-1$, where $d$ is the spatial dimension and $N$ is the number of components for the scalar field. Summations over the subscripts $\mu$ and $a$ in \Eq{eq:action} are assumed. The effective potential $V_{k}(\rho)$ is $O(N)$ invariant with $\rho=\phi^2/2$ and $\phi^2=\phi_a \phi_a$. Note that in \Eq{eq:action} we have employed the local potential approximation (LPA) supplemented with a $k$-dependent wave function renormalization $Z_{\phi,k}$, which is usually called as the truncation of $\mathrm{LPA}'$.

The evolution of effective action in \Eq{eq:action} with the RG scale is described by the Wetterich equation \cite{Wetterich:1992yh}, i.e.,
\begin{align}
  \partial_t \Gamma_{k}[\phi]&=\frac{1}{2}\mathrm{Tr}\Big \{\big(\partial_t R_k\big) G_{k}[\phi]\Big \}\,, \label{eq:WetterichEq}
\end{align}
with the RG time $t\equiv\ln (k/\Lambda)$, where $\Lambda$ is a reference scale, e.g., the UV cutoff or the initial evolution scale. The propagator in \Eq{eq:WetterichEq} reads
\begin{align}
  G_{k}[\phi]&=\frac{1}{\Gamma^{(2)}_{k}[\phi]+R_k}\,, \label{eq:Gk}
\end{align}
with 
\begin{align}
  \Gamma^{(2)}_{k}[\phi]&\equiv \frac{\delta^2 \Gamma_{k}[\phi]}{\delta \phi^2}\,. \label{}
\end{align}
Here, the bilinear regulator $R_k$ in \Eq{eq:WetterichEq} and \Eq{eq:Gk} is devised to suppress quantum fluctuations of momenta smaller than the RG scale, while make others unaltered, and see, e.g., \cite{Pawlowski:2005xe, Fu:2022gou} for more details. In this work we adopt the flat regulator \cite{Litim:2000ci,Litim:2001up}, which reads
\begin{align}
  R_{k}(q)&= Z_{\phi,k} q^2 r(q^2/k^2)\,,\label{eq:regulatorOpt}
\end{align}
with 
\begin{align}
  r(x)&= \Big(\frac{1}{x}-1\Big)\Theta(1-x)\,,\label{eq:regulatorOpt2}
\end{align}
where $\Theta(x)$ is the Heaviside step function.

The flow equation of the effective potential in \Eq{eq:action} is readily obtained from the Wetterich equation in \Eq{eq:WetterichEq}, viz.,
\begin{align}
  \partial_t V_k(\rho)=&  \mathscr{C} k^{d}\left[\frac{1}{1+\bar m_{\sigma,k}^2}+\frac{N-1}{1+\bar m_{\pi,k}^2}\right]\,, \label{eq:dtVk}
\end{align}
with the coefficient
\begin{align}
  \mathscr{C}\equiv&  \frac{1}{2}\frac{1}{(4\pi)^{d/2}}\frac{1}{\Gamma(d/2)}\left[(2-\eta)\frac{2}{d}+\eta\frac{2}{d+2}\right]
\,, \label{}
\end{align}
where the masses of the longitudinal and transversal  modes are denoted by 
\begin{align}
   \bar m_{\sigma,k}^2\equiv&\frac{1}{Z_{\phi,k} k^2}\Big(V'_k(\rho)+2 \rho V^{(2)}_k(\rho)\Big)\,,\\[2ex]
   \bar m_{\pi,k}^2\equiv&\frac{1}{Z_{\phi,k} k^2} V'_k(\rho)\,,\label{eq:masspi2}
\end{align}
respectively, and the anomalous dimension reads
\begin{align}
 \eta=&  -\frac{\partial_t Z_{\phi,k}}{Z_{\phi,k}}\,. \label{}
\end{align}

It is more convenient to work with renormalized and dimensionless variables, such that the explicit dependence of the RG scale $k$ is absorbed. To that end, one introduces
\begin{align}
 \bar \rho=&k^{-(d-2)} Z_{\phi,k} \rho\,, \qquad  u_k(\bar \rho)=k^{-d}V_k(\rho)\,.\label{}
\end{align}
Hence, the flow equation of the effective potential in \Eq{eq:dtVk} turns out to a flow of $u_k(\bar \rho)$ with $\bar \rho$ fixed, that is,
\begin{align}
 \partial_t u(\bar\rho)=&-d\, u(\bar\rho)+(d-2+\eta)\,\bar\rho \,u'(\bar\rho) \nonumber\\[2ex]
& +\mathscr{C} \left[\frac{1}{1+u'(\bar\rho)+2\bar\rho u^{(2)}(\bar\rho)}+\frac{N-1}{1+u'(\bar\rho)}\right]\,,\label{eq:dtu}
\end{align}
where the subscript $k$ for $u_k$ is not shown explicitly. In the truncation of $\mathrm{LPA}'$, the anomalous dimension can be extracted from the momentum dependence of the two-point correlation function of the transversal $\pi$ mode or the longitudinal $\sigma$ mode. Both modes give the same anomalous dimension for the Gau\ss ian fixed point, while there is indeed a difference in the case of, e.g., the Wilson-Fisher (WF) fixed point \cite{Wilson:1971dc}. In this work we adopt the anomalous dimension of the $\pi$ mode, which reads
\begin{align}
 \eta=&\frac{1}{2^{d-2}\pi^{d/2}\Gamma(1+\frac{d}{2})} \frac{\bar\rho_0 u^{(2)}(\bar\rho_0)^2}{\left[1+2\bar\rho_0 u^{(2)}(\bar\rho_0)\right]^2}\,,\label{eq:eta}
\end{align}
where $\bar\rho_0$ stands for the location of minimum of the potential $u(\bar \rho)$. The equation of fixed points for the effective potential is obtained by demanding $\partial_t u=0$, that immediately yields
\begin{align}
&(d-2+\eta)\,\bar\rho \,u'(\bar\rho)-d\, u(\bar\rho) \nonumber\\[2ex]
& +\mathscr{C} \left[\frac{1}{1+u'(\bar\rho)+2\bar\rho u^{(2)}(\bar\rho)}+\frac{N-1}{1+u'(\bar\rho)}\right]=0\,.\label{eq:fixedPointEq}
\end{align}

\section{Local solutions of the fixed-point equation}
\label{sec:local-solu}

Prior to the discussion of global solutions to the fixed-point equation in \Eq{eq:fixedPointEq}, we would like to have a brief review on two approaches that provide us with local information for the solution of effective potential, that is, the Taylor expansion at vanishing or finite field, and the Laurent expansion in terms of $1/\bar\rho$ in the limit of $\bar\rho \to \infty$.

\subsection{Taylor expansion at vanishing or finite field}
\label{subsec:Taylor-expan}

The most straightforward method to solve the flow equation of effective potential in \Eq{eq:dtu} is to expand the potential around the vanishing field $\bar\rho=0$, to wit,
\begin{align}
 u(\bar\rho)\simeq&\sum_{n=1}^{N_{\mathrm{tr}}}\frac{\lambda_n}{n!}\bar\rho^n\,,\label{eq:uTaylorZero}
\end{align}
where the field-independent term is ignored, and $N_{\mathrm{tr}}$ is the maximal order of the Taylor expansion used in a calculation. Inserting \Eq{eq:uTaylorZero} into \Eq{eq:dtu}, one is led to a set of flow equations for the expansion coefficients of effective potential, i.e.,
\begin{align}
 \partial_t \lambda_n\equiv& \beta_n(\lambda_1, \lambda_2, \cdots , \lambda_{n+1})\,,\label{eq:dtlam}
\end{align}
where the $\beta$ function of order $n$ is a function of couplings up to order of $n+1$. Evidently, the fixed points are determined by $\partial_t \lambda_n^*=0$, to wit, 
\begin{align}
 & \beta_n(\lambda_1^*, \lambda_2^*, \cdots , \lambda_{n+1}^*)=0\,,\label{eq:betaEq}
\end{align}
which constitute a set of closed equations with number $N_{\mathrm{tr}}$. Critical behaviors of flows near a fixed point is readily analyzed by linearizing the flows in \Eq{eq:dtlam} near the fixed point with $\lambda_n \simeq \lambda_n^*+\delta\lambda_n$, where $\delta\lambda_n$ is a small quantity. Hence, one arrives at
\begin{align}
 \partial_t (\delta\lambda_n)=&\sum_{n'=1}^{N_{\mathrm{tr}}} M_{n n'}\delta\lambda_{n'}\,,\label{}
\end{align}
with the stability matrix $M$ given by
\begin{align}
 M_{n n'}=&\frac{\partial \beta_n}{\partial \lambda_{n'}}\bigg|_{\lambda=\lambda^*}\,,\label{}
\end{align}
whose eigenvalues provide us with the critical exponents pertinent to the fixed point, see, e.g., \cite{Litim:2002cf} for more detailed discussions.

The Taylor expansion at vanishing field in \Eq{eq:uTaylorZero} is efficient and reliable to compute, e.g., critical exponents, when the spatial dimension is $d \to 4$, or at least not far away from $d=4$. A naive power counting immediately yields the dimension of $\lambda_n$, i.e., $[\lambda_n]=2n-(n-1)d$, if the anomalous dimension is assumed to be vanishing for the moment. Irrelevance of the expansion coefficient $\lambda_n$ demands $[\lambda_n]<0$, that leaves us with
\begin{align}
 n>&\frac{d}{d-2}\,.\label{}
\end{align}
Therefore, as the spatial dimension is approaching $d=2$ from above, the required expansion order $N_{\mathrm{tr}}$ in \Eq{eq:uTaylorZero} increases significantly, due to the rapidly increased number of relevant parameters. Generically, the convergence of \Eq{eq:uTaylorZero} becomes more and more difficult as $d \to 2$.

A natural extension of the Taylor expansion at vanishing field is to expand the effective potential at a finite field, which is able to alleviate the aforementioned problem born by the vanishing field expansion in some degree. This finite-field expansion reads
\begin{align}
 u(\bar\rho)\simeq&\sum_{n=1}^{N_{\mathrm{tr}}}\frac{\lambda_n}{n!}\big(\bar\rho-\kappa\big)^n\,,\label{eq:uTaylorFini}
\end{align}
where the expansion point $\kappa$ can be chosen to be $k$-independent or dependent. Moreover, it is convenient to choose $\kappa=\bar\rho_0$ being the minimum of the function $u(\bar\rho)$. Substituting \Eq{eq:uTaylorFini} into \Eq{eq:dtu} one is able to obtain fixed points and their relevant critical exponents by using the similar method described above, which will not be elaborated on anymore.

\subsection{Laurent expansion in the limit of large field}
\label{subsec:Laurent-expan}

When the field in \Eq{eq:uTaylorFini} is very large, e.g., $\bar\rho/\bar\rho_0\gg 1$, with $\bar\rho_0$ being the minimum of $u(\bar\rho)$, the nonlinear terms in the square bracket in \Eq{eq:fixedPointEq} can be safely neglected, and then the asymptotic behavior of $u(\bar\rho)$ in the limit of large field is readily obtained, as follows
\begin{align}
  u(\bar\rho) \sim & \,\gamma\, {\bar\rho}^{d/(d-2+\eta)}\,,\qquad \bar\rho \to \infty\,,\label{eq:uInfty}
\end{align}
with a constant $\gamma$. The missing subleading terms on the right side of \Eq{eq:uInfty} can be formulated with a Laurent expansion in terms of $1/\bar\rho$. Firstly, let us consider the case that the leading power $d/(d-2+\eta)$ is an integer \cite{Litim:2016hlb, Juttner:2017cpr}, and a more general case is discussed in \sec{subsubsec:Laurent-expan-general}. The expansion in the limit of large field reads
\begin{align}
  u(\bar\rho)\simeq& \,\gamma\, {\bar\rho}^{d/(d-2+\eta)}\left[1+\sum_{n=1}^{N_{\mathrm{tr}}} \gamma_{n}\left(\frac{1}{\bar\rho}\right)^n\right]\,.\label{eq:uLaurent}
\end{align}
In the same way, the expansion coefficients $\gamma_{n}$ can be extracted by inserting \Eq{eq:uLaurent} into the fixed-point equation \labelcref{eq:fixedPointEq}. In the following, we present the first few nonvanishing coefficients for some values of $d$ with $\eta=0$. When $d=3$, one is left with
\begin{align}
    \gamma_5=&\mathscr{C}\frac{5N-4}{75\gamma^2}\,,\quad \gamma_7=-\mathscr{C}\frac{25N-24}{1575\gamma^3}\,,\nonumber\\[2ex]
    \gamma_9=&\mathscr{C}\frac{125N-124}{30375\gamma^4}\,, \cdots \,. \label{eq:Laurentd3}
\end{align}
Here, the order of the first nonvanishing coefficient is $n=5$. In the case of $d=2.5$, one arrives at
\begin{align}
    \gamma_9=&2\mathscr{C}\frac{9N-8}{405\gamma^2}\,,\quad\gamma_{13}=-2\mathscr{C}\frac{81N-80}{26325\gamma^3}\,,\nonumber\\[2ex]
    \gamma_{17}=&2\mathscr{C}\frac{729N-728}{1549125\gamma^4}\,, \cdots \,.\label{}
\end{align}
For $d=2.1$, one has
\begin{align}
    \gamma_{41}=&10\mathscr{C}\frac{41N-40}{35301\gamma^2}\,,\quad\gamma_{61}=-10\mathscr{C}\frac{1681N-1680}{45220581\gamma^3}\,,\nonumber\\[2ex]
    \gamma_{81}=&10\mathscr{C}\frac{68921N-68920}{51700467861\gamma^4}\,, \cdots \,.\label{}
\end{align}
Evidently, as the spatial dimension $d$ decreases towards $d=2$, the order of the first nonvanishing coefficient increases significantly.

\subsubsection{Laurent expansion in the limit of large field for the case of $d/(d-2+\eta)$ being a rational fraction}
\label{subsubsec:Laurent-expan-general}

We have discussed above the Laurent expansion of potential in the limit of large field, cf. \Eq{eq:uLaurent}, where the leading power $d/(d-2+\eta)$ is an integer. In this subsection, we discuss a more general case with $d/(d-2+\eta)$ being a rational fraction. Note that even it is an irrational number, one can always approximate it with a rational one up to any desired accuracy. Let the leading power be
\begin{align}
 \frac{d}{d-2+\eta}=&\frac{l}{m}\,,\label{eq:fraclm}
\end{align}
where the right side has been fully simplified, and  both $l$ and $m$ are integers. Then the Laurent expansion in \Eq{eq:uLaurent} can be modified as
\begin{align}
    u(\bar\rho)\simeq& \,\gamma\, {\bar\rho}^{d/(d-2+\eta)}\left[1+\sum_{n=1}^{N_{\mathrm{tr}}} \gamma_{n}\left(\frac{1}{\bar\rho^{\frac{1}{m}}}\right)^n\right]\,.\label{eq:uLaurent2}
\end{align}

Here we show some examples with $\eta=0$ and the dimension in the vicinity of $d=3$ . Firstly, considering the case of $d=2.9$, one has $d/(d-2+\eta)=29/9$. In the same way, inserting \Eq{eq:uLaurent2} into the fixed-point equation \labelcref{eq:fixedPointEq}, one is able to obtain the first few nonvanishing coefficients, as follows
\begin{align}
    \gamma_{49}=&90\mathscr{C}\frac{49N-40}{69629\gamma^2}\,,\quad \gamma_{69}=-270\mathscr{C}\frac{2401N-2320}{46442543\gamma^3}\,,\nonumber\\[2ex]
    \gamma_{89}=&7290\mathscr{C}\frac{117649N-116920}{255371390029\gamma^4}\,, \cdots \,. \label{}
\end{align}
Then the potential in \Eq{eq:uLaurent2} reads
\begin{align}
    u(\bar\rho)\Big|_{d=2.9}\simeq& \,\gamma\, {\bar\rho}^{3.22}+\gamma \gamma_{49}{\bar\rho}^{-2.22}+\gamma \gamma_{69}{\bar\rho}^{-4.44}\nonumber\\[2ex]
&+\gamma \gamma_{89}{\bar\rho}^{-6.67}+\cdots\,.\label{eq:ud2d9}
\end{align}
If the value of $d$ is increased up to $d=2.96$, the fraction in \Eq{eq:fraclm} reads $l/m=37/12$. It follows that
\begin{align}
    \gamma_{62}=&75\mathscr{C}\frac{31N-25}{35557\gamma^2}\,,\quad \gamma_{87}=-600\mathscr{C}\frac{961N-925}{38152661\gamma^3}\,,\nonumber\\[2ex]
    \gamma_{112}=&1350\mathscr{C}\frac{29791N-29575}{10563024661\gamma^4}\,, \cdots \,, \label{}
\end{align}
which yields
\begin{align}
    u(\bar\rho)\Big|_{d=2.96}\simeq& \,\gamma\, {\bar\rho}^{3.08}+\gamma \gamma_{62}{\bar\rho}^{-2.08}+\gamma \gamma_{87}{\bar\rho}^{-4.17}\nonumber\\[2ex]
&+\gamma \gamma_{112}{\bar\rho}^{-6.25}+\cdots\,.\label{eq:ud2d96}
\end{align}
From \Eq{eq:Laurentd3}, one immediately finds for the Laurent expansion of potential with $d=3$
\begin{align}
    u(\bar\rho)\Big|_{d=3}\simeq& \,\gamma\, {\bar\rho}^{3}+\gamma \gamma_{5}{\bar\rho}^{-2}+\gamma \gamma_{7}{\bar\rho}^{-4}+\gamma \gamma_{9}{\bar\rho}^{-6}+\cdots\,.\label{eq:ud3}
\end{align}
From Eqs. \labelcref{eq:ud2d9}, \labelcref{eq:ud2d96}, \labelcref{eq:ud3}, one immediately finds that as the dimension $d$ approaches towards $d=3$, the powers of $\bar \rho$ of different orders converges at their respective values of $d=3$.

\section{Global solutions of the fixed-point equation}
\label{sec:global-solu}

In this section we would like to solve the fixed-point equation in \Eq{eq:fixedPointEq} directly by using numerical method. \Cref{eq:fixedPointEq} is a differential algebraic equation (DAE) of index 1 \cite{Campbell:2008}, which indicates that one has to take one further derivative on the DAE to transform it into a set of ordinary differential equations (ODE), and see e.g., \cite{Grossi:2019urj} for relevant discussions. The resulting ODEs read
\begin{subequations}
    \begin{align}
    u'(\bar\rho)=&u_1(\bar\rho)\,,\label{eq:dudrho1}\\[2ex]
    u_1'(\bar\rho)=&u_2(\bar\rho)\,,\label{eq:dudrho2}\\[2ex]
    u_2'(\bar\rho)=&\frac{1}{2\mathscr{C}\bar\rho}\Big(1+u_1(\bar\rho)+2\bar\rho u_2(\bar\rho)\Big)^2\bigg[(-2+\eta)u_1(\bar\rho)\nonumber\\[2ex]
&+(d-2+\eta)\bar\rho u_2(\bar\rho)-\mathscr{C}(N-1)\frac{u_2(\bar\rho)}{\big(1+u_1(\bar\rho)\big)^2}\bigg]\nonumber\\[2ex]
&-\frac{3}{2\bar\rho}u_2(\bar\rho)\,.\label{eq:dudrho3}
    \end{align}
    \label{eq:dudrhos}
\end{subequations}
Here, we have defined $u_1$ and  $u_2$ in \labelcref{eq:dudrho1} and \labelcref{eq:dudrho2}, which correspond to the first and second derivatives of $u(\bar\rho)$, respectively. \Cref{eq:dudrho3} is obtained by differentiating \Eq{eq:fixedPointEq} with respect to $\bar\rho$.

We integrate the differential equations in \labelcref{eq:dudrhos}, starting at a sufficient large $\bar\rho=\bar\rho_{L}$ with initial values for $u(\bar\rho_{L})$, $u_1(\bar\rho_{L})$ and $u_2(\bar\rho_{L})$ obtained from the asymptotic expression of $u(\bar\rho)$ in the limit of large field as shown in \Eq{eq:uInfty}. The anomalous dimension $\eta$ in \labelcref{eq:dudrhos} and  \labelcref{eq:uInfty} is given an initial value, say $\eta=0$. The functions are integrated out from $\bar\rho=\bar\rho_{L}$ towards $\bar\rho \to 0$. The parameter $\gamma$ in \Eq{eq:uInfty} is fine tuned such that the potential $u(\bar\rho)$ and its derivatives of different orders $u^{(n)}(\bar\rho)$ are finite in the limit of $\bar\rho \to 0$. Then, the obtained potential is inserted back into \Eq{eq:eta} to update the value of $\eta$. Iterate the process mentioned above until convergence is obtained. Using such approach, one is able to find a solution to the fixed-point equation \labelcref{eq:fixedPointEq}, denoted by $u^{*}(\bar\rho)$ in what follows.

%
\begin{figure*}[t]
\includegraphics[width=0.45\textwidth]{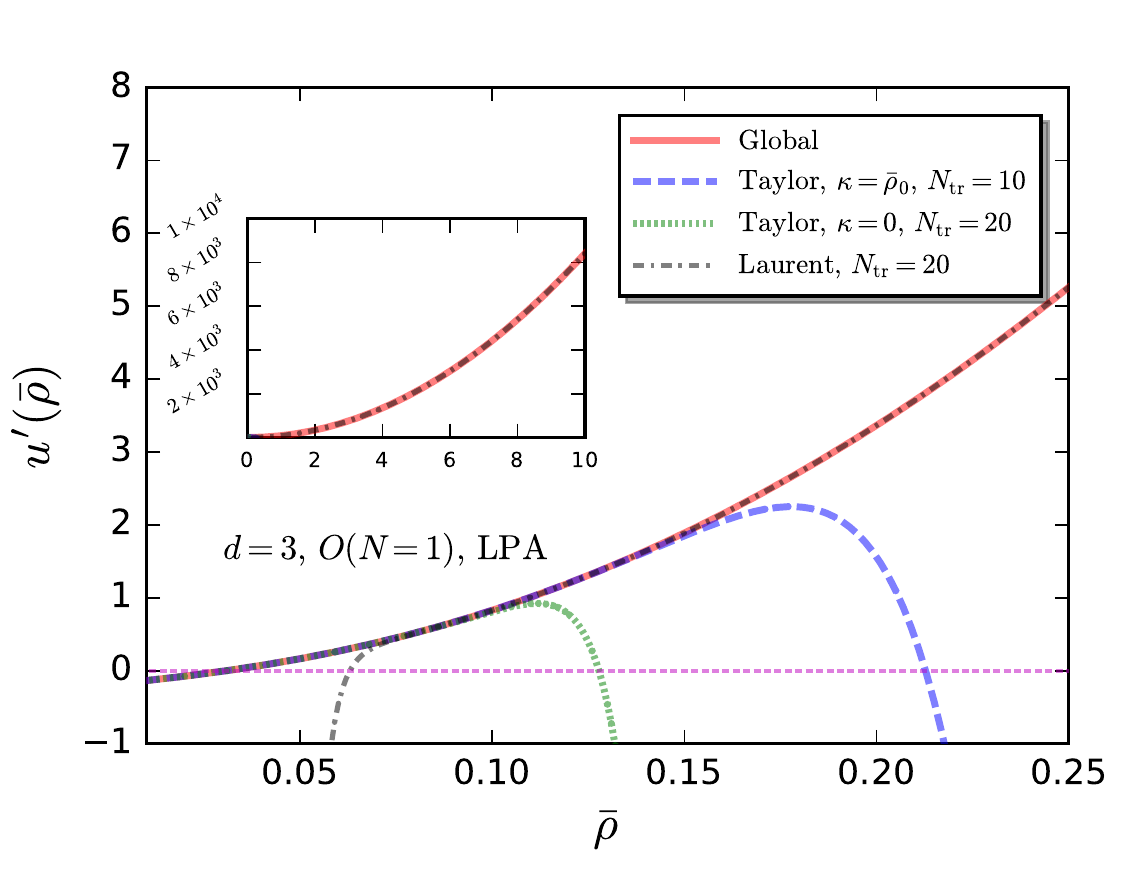}\hspace{0.5cm}
\includegraphics[width=0.45\textwidth]{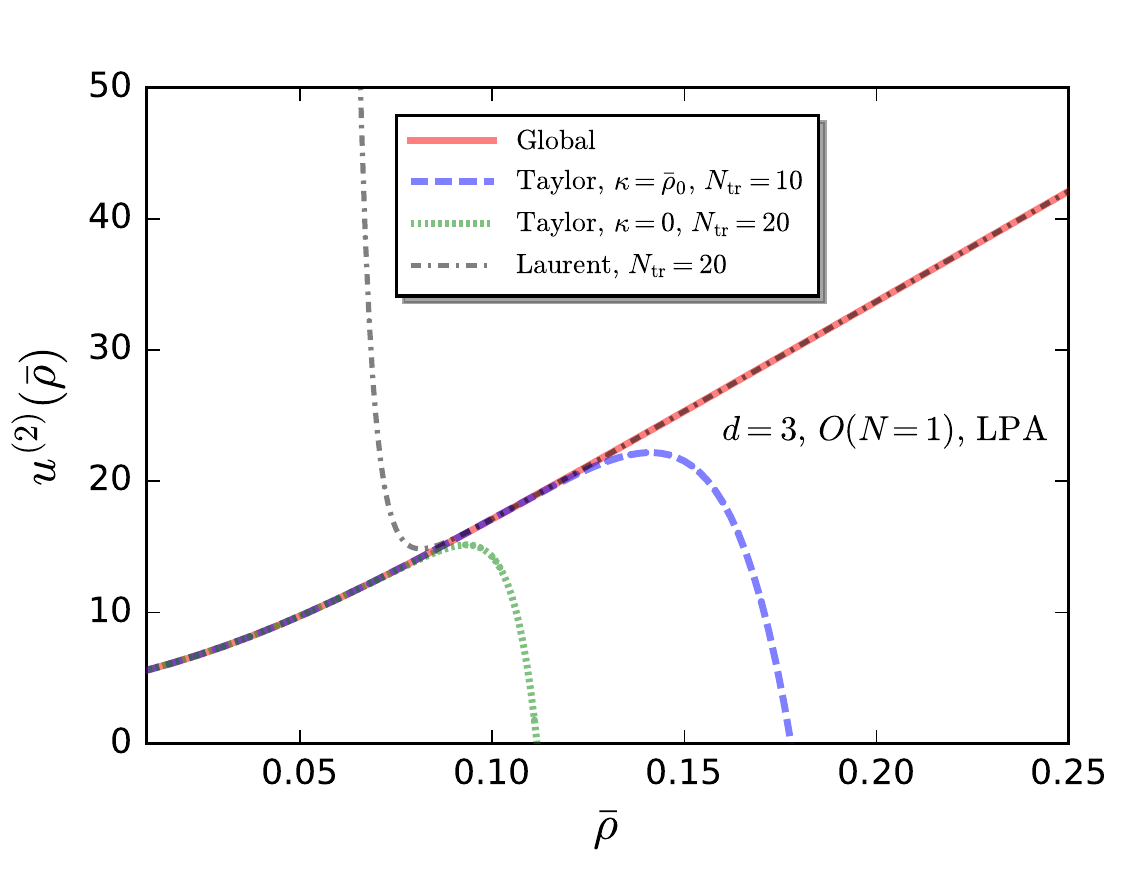}
\caption{Derivatives of the effective potential $u^\prime(\bar\rho)$ (left panel) and $u^{(2)}(\bar\rho)$ (right panel) at the Wilson-Fisher fixed point as functions of $\bar\rho$ for the $d=3$ dimensional $O(1)$ scalar theory, i.e., the Ising universality class, obtained in the truncation of LPA. The global potential is in comparison to those obtained from Taylor and Laurent expansions, and for the Taylor expansion both the vanishing and finite expansion points are employed. The zoomed-out view of $u^\prime(\bar\rho)$ is also shown in the inlay of the left panel.}\label{fig:du-N1-d3}
\end{figure*}
%

%
\begin{figure}[t]
\includegraphics[width=0.45\textwidth]{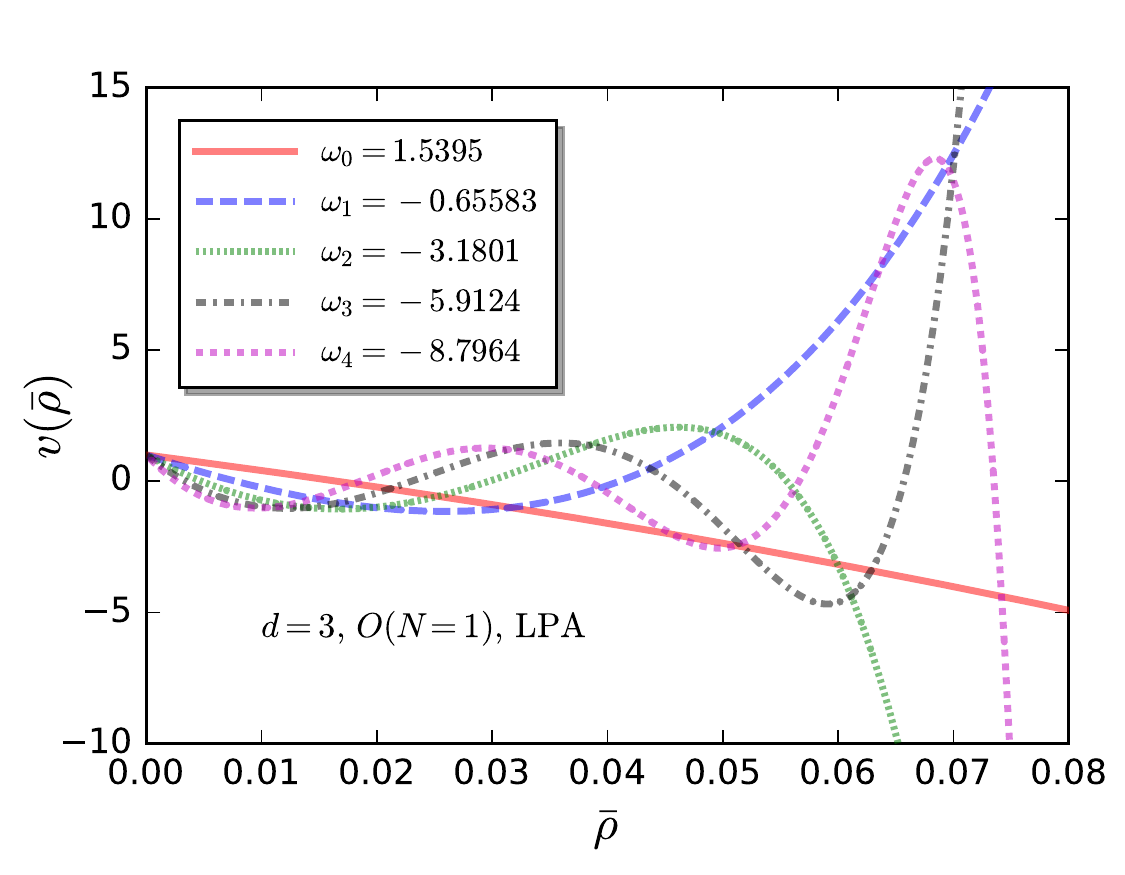}
\caption{Eigenfunctions of the first few eigenvalues for the potential of WF fixed point shown in \Fig{fig:du-N1-d3}, where we have used the normalization $v(0)=1$ and the calculation is done for the $3d$ $O(1)$ scalar theory in LPA approximation.}\label{fig:v-N1-d3}
\end{figure}
%

%
\begin{figure*}[t]
\includegraphics[width=0.45\textwidth]{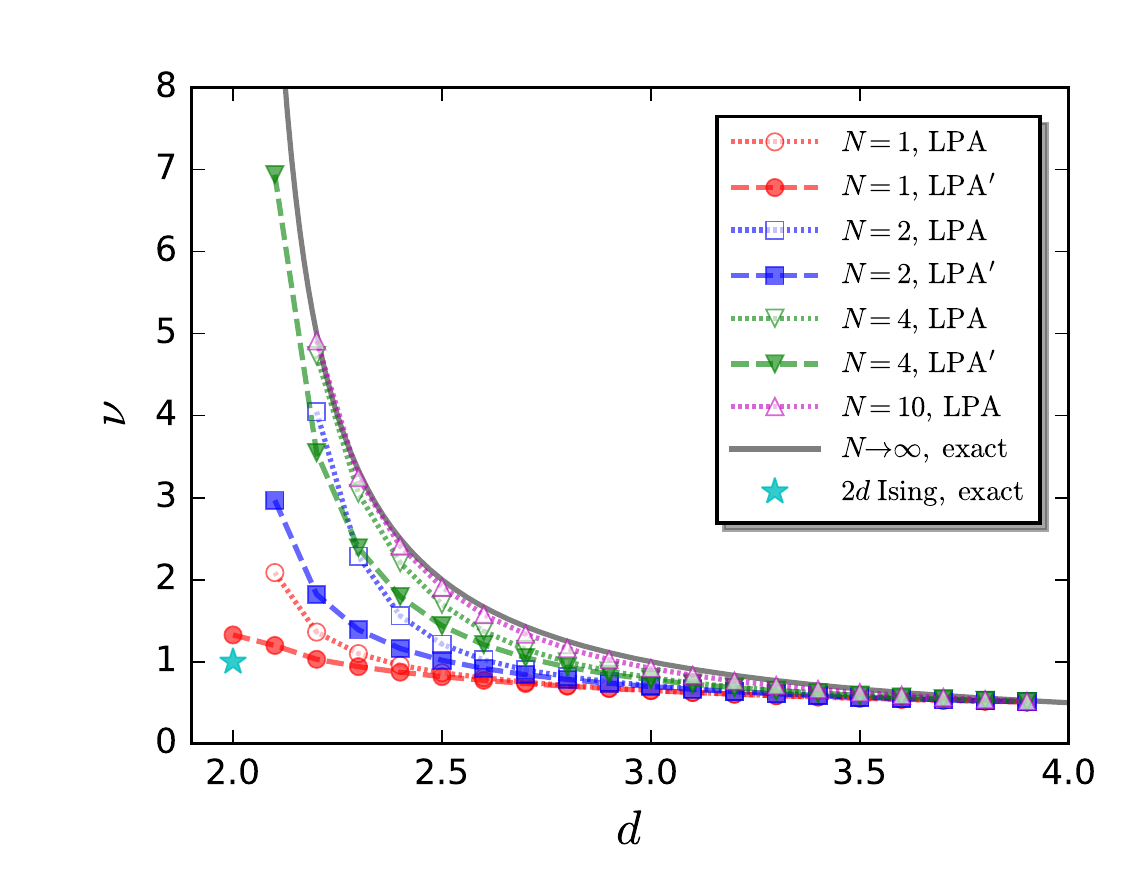}\hspace{0.5cm}
\includegraphics[width=0.45\textwidth]{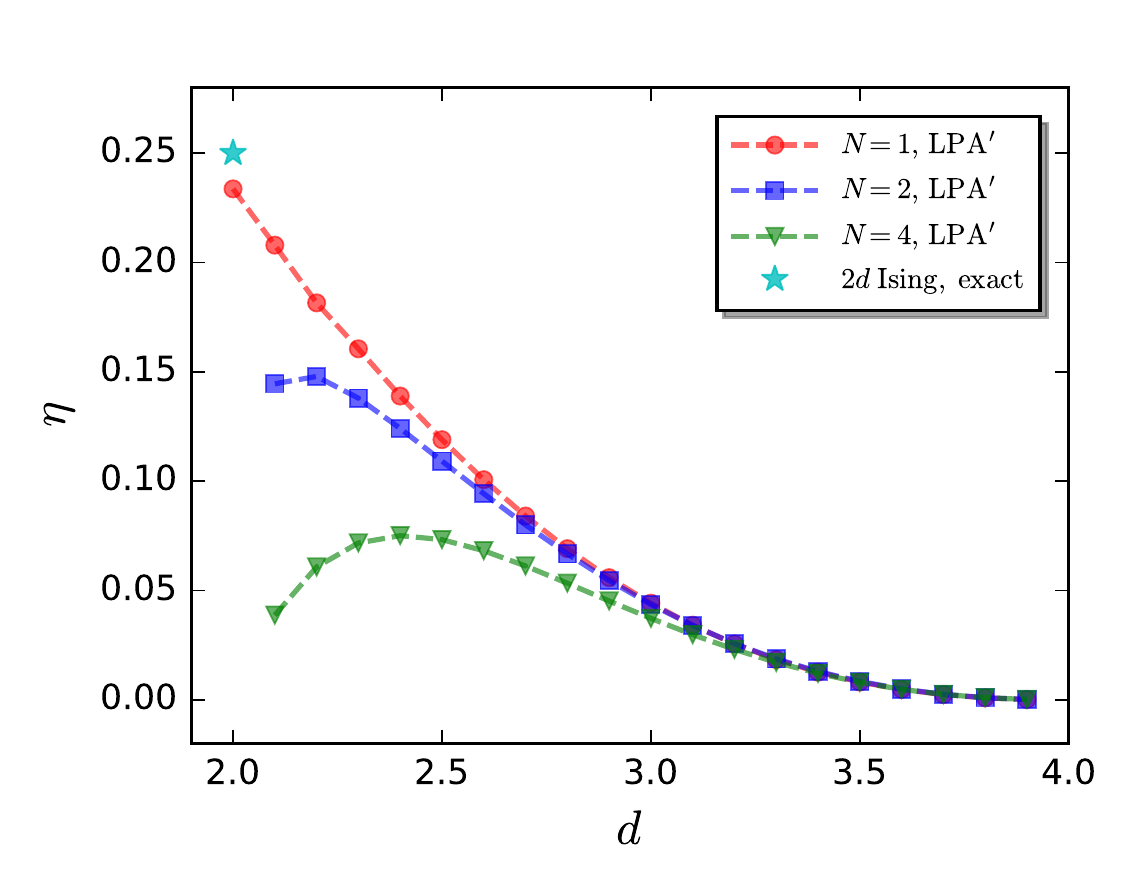}
\caption{Critical exponents $\nu$ (left panel) and $\eta$ (right panel) of the $O(N)$ universality class as functions of the spatial dimension $d$ with several different values of $N$ obtained in LPA and $\mathrm{LPA}'$, where $d$ is a continuous variable in the range $2\leq d \leq 4$. The exact results for the $2d$ Ising model and the spherical model with $N \to \infty$ are also shown.}\label{fig:nu-eta}
\end{figure*}
%

%
\begin{figure}[t]
\includegraphics[width=0.45\textwidth]{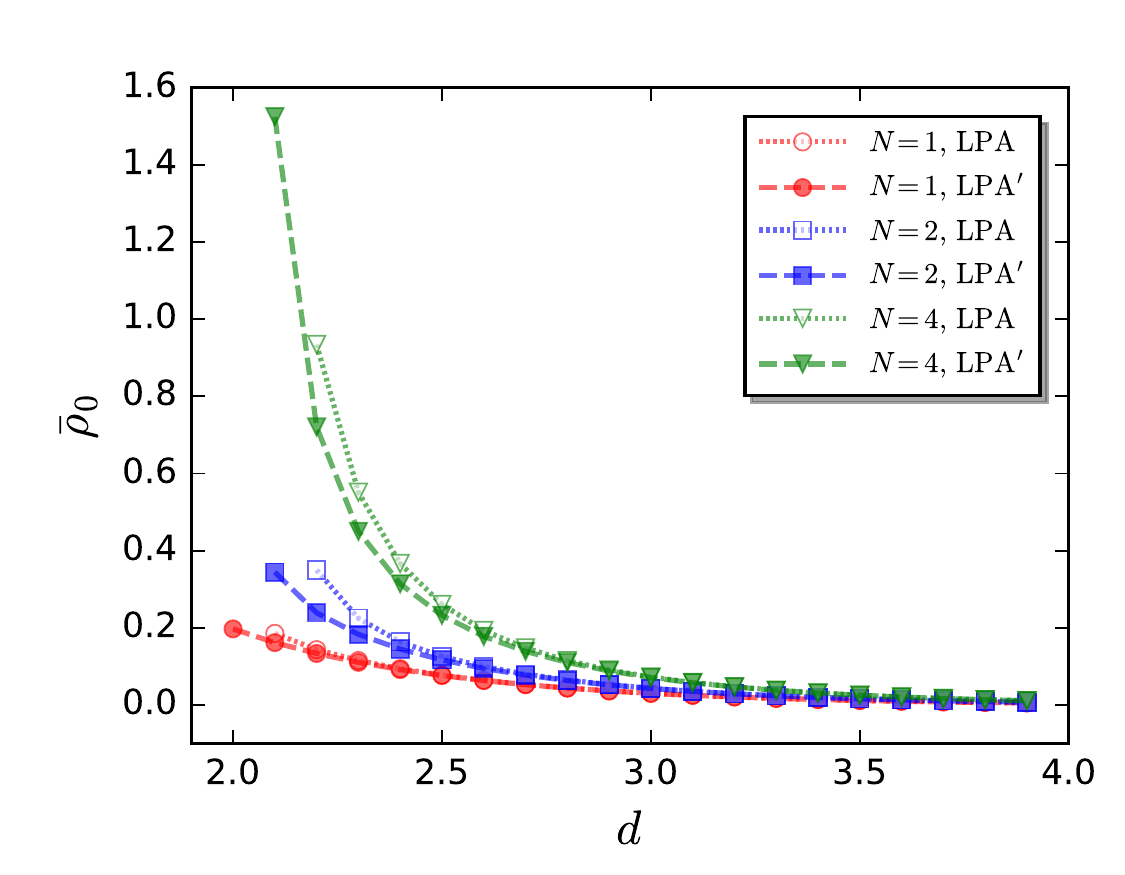}
\caption{Location of the minimum of the potential $u(\bar\rho)$ at the Wilson-Fisher fixed point, $\bar \rho_0$, as a function of the spatial dimension $d$ with several different values of $N$ obtained in the truncations of LPA and $\mathrm{LPA}'$.}\label{fig:rho0}
\end{figure}
%

We proceed with the discussion of eigenperturbations near the fixed-point potential $u_{*}(\bar\rho)$ and the relevant eigenvalues, i.e., critical exponents \cite{Codello:2014yfa}. The effective potential near the fixed point can be written as
\begin{align}
 u(\bar\rho)=&u_{*}(\bar\rho)+\epsilon\,\mathrm{e}^{-\omega t}v(\bar\rho)\,,\label{eq:eigenpertur}
\end{align}
where $\epsilon$ is a small parameter, $v(\bar\rho)$ and $\omega$ are the eigenfunction and the corresponding eigenvalue. One can see that only perturbations of eigenvalue $\omega>0$ are relevant. Inserting \Eq{eq:eigenpertur} into the flow equation of the effective potential in \Eq{eq:dtu} and reformulating the equation by powers of the small parameter $\epsilon$, one immediately finds that the fixed-point equation in \Eq{eq:fixedPointEq} is reproduced in the zeroth-order term, and the linear term in $\epsilon$ provides us with a homogeneous differential equation for the eigenfunction $v(\bar\rho)$ that we are looking for, i.e.,
\begin{align}
  \omega v(\bar\rho)=&d v(\bar\rho)-(d-2+\eta)\bar\rho v'(\bar\rho)\nonumber\\[2ex]
&+\mathscr{C}\Bigg[\frac{v'(\bar\rho)+2\bar\rho v^{(2)}(\bar\rho)}{\big(1+u_*^{\prime}(\bar\rho)+2\bar\rho u_*^{(2)}(\bar\rho)\big)^2}+\frac{(N-1)v'(\bar\rho)}{\big(1+u_*^{\prime}(\bar\rho)\big)^2}\Bigg]\,.\label{eq:dvdrho}
\end{align}
The asymptotic behavior of $v(\bar\rho)$ in the limit of $\bar\rho\to\infty$ is readily obtained from \Eq{eq:dvdrho} by ignoring the subleading terms in the square bracket, since they are suppressed by $u_*^{\prime}(\bar\rho)$ and $u_*^{(2)}(\bar\rho)$ in the denominator. Thus, one is led to
\begin{align}
  v(\bar\rho) \simeq&\,c \,{\bar\rho}^{(d-\omega)/(d-2+\eta)}\,,\qquad \bar\rho \to \infty\,.\label{eq:vInfty}
\end{align}
with a normalized constant $c$. In the same way, as $\bar\rho \to 0$ one obtains from \Eq{eq:dvdrho} the relation as follows
\begin{align}
  v'(0) =&\frac{1}{\mathscr{C} N}(\omega-d)\big(1+u_*^{\prime}(0)\big)^2v(0)\,.\label{eq:v0}
\end{align}

Similarly as solving the equations in \labelcref{eq:dudrhos}, we integrate the differential equation for the eigenfunction $v(\bar\rho)$ in \Eq{eq:dvdrho}, starting at a sufficient large value of $\bar\rho$, e.g., $\bar\rho=\bar\rho_{L}$ with initial values for $v(\bar\rho_{L})$ and $v'(\bar\rho_{L})$ obtained in \Eq{eq:vInfty}. The function $v(\bar\rho)$ is resolved as the differential equation is integrated out from $\bar\rho=\bar\rho_{L}$ to $\bar\rho \to 0$. We fine tune the value of $\omega$, such that $v(\bar\rho)$ and $v'(\bar\rho)$ are finite in the limit of $\bar\rho \to 0$, or more exactly the relation in \Eq{eq:v0} is satisfied. Then one can obtain the eigenfunction $v(\bar\rho)$ and its related eigenvalue $\omega$.

\section{Numerical results}
\label{sec:num-resul}

%
\begin{table*}[t]
  \begin{center}
  \begin{tabular}{cc|cc|cc}
     \hline\hline & & & & &  \\
     & Method & \multicolumn{2}{c|}{$d=3$} & \multicolumn{2}{c}{$d=2$}\\[2ex]
     \hline
    & & $\nu$ &  $\eta$  & $\nu$ &  $\eta$  \\[1ex]
    \hline & & & & &  \\[-2ex]
    $O(1)$ LPA (this work) & fRG global (direct) &0.6495619 &0 \\[1ex]
    $O(1)$ LPA$'$ (this work) & fRG global (direct) &0.6473203 &0.0442723  &1.3266022 &0.2335624 \\[1ex]
    $O(4)$ LPA (this work) & fRG global (direct) &0.8043477 &0 \\[1ex]
    $O(4)$ LPA$'$ (this work) & fRG global (direct) &0.7811038 &0.0373204 \\[1ex]
    $O(40)$ LPA (this work) & fRG global (direct) &0.9807813 &0 \\[1ex]
    $O(1)$ LPA \cite{Juttner:2017cpr}  & fRG global (combination) &0.6495618 &0 \\[1ex]
    $O(4)$ LPA \cite{Juttner:2017cpr} & fRG global (combination) &0.8043477 &0 \\[1ex]
    $O(40)$ LPA \cite{Juttner:2017cpr} & fRG global (combination) &0.9807813 &0 \\[1ex]
    $O(1)$ LPA$'$ \cite{Borchardt:2015rxa}& fRG global (pseudospectral) &0.645995 &0.0442723  & & \\[1ex]
    $O(1)$ LPA$'$ \cite{Codello:2012sc, Codello:2012ec, Codello:2014yfa} & fRG iterative &0.65 &0.044  &1.33 &0.23 \\[1ex]
    $O(4)$ LPA$'$ \cite{Codello:2012sc, Codello:2012ec, Codello:2014yfa} & fRG iterative &0.78 &0.037  & & \\[1ex]
    $O(1)$ scalar theories \cite{Balog:2019rrg, DePolsi:2020pjk} & fRG DE $\mathcal{O}(\partial^6)$ & 0.63012(5) & 0.0361(3) &\\[1ex]
    $O(4)$ scalar theories \cite{DePolsi:2020pjk} & fRG DE $\mathcal{O}(\partial^4)$ & 0.7478(9) & 0.0360(12) &\\[1ex]
    $O(1)$ CFTs \cite{Kos:2014bka} & conformal bootstrap & 0.629971(4) & 0.0362978(20) &\\[1ex]
    $O(4)$ CFTs \cite{Kos:2015mba} & conformal bootstrap & 0.7472(87) & 0.0378(32) &\\[1ex]
    $O(4)$ spin model \cite{Kanaya:1994qe}  &Monte Carlo & 0.7479(90) & 0.025(24)  &\\[1ex]
    $O(1)$ (Ising)   & exact &          & &1 &1/4         \\[1ex]
    $O(N \to \infty)$    &exact &$1/(d-2)$ &  0        &$1/(d-2)$ &0         \\[1ex]
    \hline\hline
  \end{tabular}
  \caption{Critical exponents $\nu$ and $\eta$ of the $O(N)$ scalar theory in $d=3$ and 2 spatial dimension obtained in fRG with truncations LPA and LPA$'$, where several different values of $N$ are adopted. Note that the $O(1)$ symmetry corresponds to that of the Ising model. Our results are in comparison to those from other approaches, e.g., the global fixed points with a combination of analytical and numerical techniques \cite{Juttner:2017cpr}, the pseudospectral methods \cite{Borchardt:2015rxa}, the iterative method \cite{Codello:2012ec, Codello:2012sc, Codello:2014yfa}, derivative expansions \cite{Balog:2019rrg, DePolsi:2020pjk}, the conformal bootstrap for the $3d$ conformal field theories (CFTs) \cite{Kos:2014bka, Kos:2015mba}, Monte Carlo simulations \cite{Kanaya:1994qe}. Moreover, exact results for the $2d$ Ising model and the $O(N)$ symmetry with $N \to \infty$ are also presented.} 
  \label{tab:exponent}
  \end{center}\vspace{-0.5cm}
\end{table*}
%

In this work we employ the fifth order Radau IIA method (RadauIIA5) \cite{Hairer:1996,Hairer:1999} encoded in a \textit{Julia} package \cite{Rackauckas:2017} to solve the differential equations in \labelcref{eq:dudrhos} numerically. 

In \Fig{fig:du-N1-d3} we show the first and second derivatives of the effective potential  $u^\prime(\bar\rho)$ and $u^{(2)}(\bar\rho)$ at the Wilson-Fisher fixed point for the scalar theory of the $O(1)$ symmetry in $d=3$ spatial dimension, obtained in the LPA approximation. As we have discussed above, the global potential is obtained by integrating the differential equations in \Eq{eq:dudrhos} from a sufficient large $\bar\rho=\bar\rho_{L}$ towards $\bar\rho \to 0$. Usually the value of $\bar\rho_{L}$ should be chosen large enough such that one has $\bar\rho_{L}\gg \bar\rho_0$, where $\bar\rho_0$ is the location of the minimum of potential $u(\bar\rho)$, and the obtained potential would not show dependence on the choice of $\bar\rho_{L}$. It is found that $\bar\rho_{L}=10$ would meet the requirements in \Fig{fig:du-N1-d3}. The leading order expansion coefficient of the potential at large field, i.e., $\gamma$ in \Eq{eq:uInfty} or \Eq{eq:uLaurent}, is fine tuned to pin down the desired solution of fixed point as discussed in \sec{sec:global-solu}. For the WF fixed point, we find $\gamma=28.060767758247646$ that is in good agreement with $\gamma=28.060767757778700$ obtained from pseudospectral methods \cite{Borchardt:2015rxa}, being identical for the first 10 significant digits. Moreover, the location of minimum of the potential, i.e., the crossing point between the curve of $u^\prime$ and the horizontal zero dashed line as shown in the left panel of \Fig{fig:du-N1-d3}, is found to be $\bar\rho_0=0.03064794240852456$, which is also consistent with $\bar\rho_0=0.03064794240869777$ from pseudospectral methods \cite{Borchardt:2015rxa}. There are several sources for the numerical errors in our calculations, for instance, numerical solving of the fixed-point equations in \labelcref{eq:dudrhos} and the eigenfunction equation in \labelcref{eq:dvdrho}, iterative solution of the anomalous dimension in the case of $\mathrm{LPA}'$. The numerical errors are in fact only restricted by the accuracy of computing machine, and thus can be neglected in comparison to the systematic errors resulting from the derivative expansion of the effective action.

In \Fig{fig:du-N1-d3} the global solution of potential is also compared with local solutions obtained from the Taylor expansion with expansion points $\kappa=0$ and $\kappa=\bar\rho_0$ and from the Laurent expansion in the limit of $\bar\rho \to \infty$. The maximal order of expansion $N_{\mathrm{tr}}$ is 10 for the Taylor expansion with $\kappa=\bar\rho_0$ and 20 for the two others. One can see that the global $u^\prime$ coincides with that of the Taylor expansion in the regime of small $\bar\rho$, and an obvious deviation takes place at about $\bar\rho\simeq 0.1$ for $\kappa=0$ and $\bar\rho\simeq 0.15$ for $\kappa=\bar\rho_0$. This also indicates that Taylor expansion at finite field is superior to that at vanishing field, though the expansion order $N_{\mathrm{tr}}$ of the former is smaller. On the contrary, the deviation between the global solution and the Laurent expansion happens in the region of small field, say $\bar\rho\lesssim 0.07$ as shown in the left panel of \Fig{fig:du-N1-d3}, but they are in good agreement when the field is large. Similar behaviors are also found in $u^{(2)}$ as shown in the right panel of \Fig{fig:du-N1-d3}.

Eigenfunctions of the first several low orders and their respective eigenvalues related to the potential of the WF fixed point in \Fig{fig:du-N1-d3} are presented in \Fig{fig:v-N1-d3}. Here, the different values of the eigenvalue $\omega$ in \Eq{eq:eigenpertur} are denoted by $\omega_n$, with $n$ standing for the order of the eigenfunction. Since the WF fixed point is a stable fixed point, this indicates that there is only one positive eigenvalue with $\omega_0=1.5395$, that is, only one mode of eigenperturbation is relevant for the WF fixed point. This eigenvalue allows us to obtain the critical exponent $\nu=1/\omega_0=0.64956$ for the $3d$ Ising model in LPA approximation.

In \Tab{tab:exponent} our calculated critical exponents $\nu$ and $\eta$ in $d=3$ and 2 spatial dimension  with different values of $N$ for the $O(N)$ symmetry universality class are shown, which are also compared with relevant results from other approaches in the literatures. For instance, the values of $\nu$ in LPA with $d=3$ obtained in this work are in good agreement with those obtained from the same truncation in \cite{Juttner:2017cpr}, being identical for the first $6\sim7$ significant digits. Note that rather than a direct solution of the fixed point equation, a quite different approach, that is, a combination of analytical and numerical techniques, is used to determine global fixed points in \cite{Juttner:2017cpr}. The critical exponents for the $2d$ Ising model are also calculated in $\mathrm{LPA}'$ with $\nu=1.327$ and $\eta=0.2336$, which are consistent with the results obtained from the iterative method \cite{Codello:2012ec, Codello:2012sc, Codello:2014yfa}, and comparable to the exact results of $\nu=1$ and $\eta=0.25$. In order to make a comparison to the exact results of $\nu=1/(d-2)$ and $\eta=0$ in the limit of $N \to \infty$, i.e., the critical exponents in the spherical model. We have done a computation with $N=40$ and $d=3$ in LPA and found $\nu=0.9808$, being very close to the exact result.

In \Fig{fig:nu-eta} the dependence of the critical exponents $\nu$ and $\eta$ for the $O(N)$ universality class on the spatial dimension $d$ and the number of field components $N$ is investigated. Here the dimension $d$ is not constrained to be an integer anymore, but rather a continuous variable in the range $2\leq d \leq 4$. Both truncations LPA and $\mathrm{LPA}'$ are used. The exact results for the $2d$ Ising model and the limit $N \to \infty$ are presented for comparison. One can see when $d \to 4$, the Wilson-Fisher fixed point coincides with the Gau\ss ian fixed point, and all the curves converge at the trivial results of $\nu=1/2$ and $\eta=0$. With the decrease of $d$ from 4 to 2, $\nu$ increases for each value of $N$, and the magnitude of increase is larger for larger $N$, but this behavior is rapidly saturated around $\nu=1/(d-2)$ in the limit of $N \to \infty$.  $\nu=1.327$ and $\eta=0.2336$ is found for $d=2$ and $N=1$ in $\mathrm{LPA}'$. We find that the calculations for $N \geq 2$ in $d=2$ are quite difficult. This can also be inferred from the asymptotic behavior of the potential in the limit of large field in \Eq{eq:uInfty}, where the power $d/(d-2+\eta)$ is divergent when $d=2$ and $\eta=0$. In fact, this behavior is closed related to the Mermin-Wagner-Hohenberg theorem \cite{Mermin:1966fe, Hohenberg:1967zz, Coleman:1973ci}, stating that the continuous symmetries, e.g., $N \geq 2$ for the $O(N)$ universality class, would not be broken spontaneously in $d=2$ dimension, that indicates one has $\nu \to \infty$ for $N \geq 2$ in $d=2$ dimension. Moreover, it is found that the anomalous dimension $\eta$ increases monotonically with the decrease of $d$ from 4 to 2 for $N=1$, while it is not monotonic for $N\geq 2$. In \Fig{fig:rho0} we show a non-universal variable, the location of minimum of the potential $u(\bar\rho)$ at the Wilson-Fisher fixed point, i.e., $\bar \rho_0$, and its dependence on the dimension $d$ with several different values of $N$. One can see that $\bar \rho_0$ also increases rapidly for $N \geq 2$ when $d$ is approaching 2.

\section{Summary and outlook}
\label{sec:summary}

In this work the fixed-point equation for the nonperturbative effective potential in the fRG approach is integrated from large to vanishing field, where the asymptotic potential in the limit of large field is implemented as initial conditions. This approach provides us with a global fixed-point potential with high numerical accuracy, that captures both the asymptotic behavior in the limit of vanishing field and more importantly that in the limit of large field. The obtained global potential is in good agreement with the results from the Taylor expansion and the Laurent expansion in the regimes where they are applicable, i.e., small field for the former and large field for the latter, respectively. Furthermore, Laurent expansion of the potential in the limit of large field for the general case, that the spatial dimension $d$ is a continuous variable in the range $2\leq d \leq 4$, is obtained.

By virtue of the method of eigenperturbations, we also compute the eigenfunctions and eigenvalues of perturbations near the Wilson-Fisher fixed point with high numerical accuracy. Consequently, critical exponents for different values of the spatial dimension $d$ and the number of field components $N$ of the $O(N)$ universality class are obtained. Our calculated critical exponents are in good agreement with the relevant results in the literatures with the same truncation, and are also comparable with the exact results for the $2d$ Ising model and the spherical model with $N \to \infty$.

Furthermore, it is also desirable to apply the approach used in this work to other physical problems of interest, such as the dynamical critical exponent \cite{Tan:2021zid}, the Yang--Lee edge singularity \cite{Stephanov:2006dn, Mukherjee:2019eou, Connelly:2020gwa, Rennecke:2022ohx, Ihssen:2022xjv}, multi-critical fixed points \cite{Yabunaka:2017uox}, etc. Extension of this approach to higher-order derivative expansions of effective action beyond $\mathrm{LPA}'$ is also very valuable.


\begin{acknowledgments}
We thank Jan M. Pawlowski, Nicolas Wink for discussions. This work is supported by the National Natural Science Foundation of China under Grant No. 12175030. 
\end{acknowledgments}


	
\bibliography{ref-lib}

\end{document}